\documentclass[a4paper,11pt]{article}
\usepackage{pos}

\usepackage{subfig}

\begin{document}
\title{High-energy reconstruction for single and double cascades using the KM3NeT detector}

\author*[a]{Thijs van Eeden}
\author*[a]{Jordan Seneca}
\author[a]{Aart Heijboer}
\affiliation[a]{Nikhef,\\ Science Park 105, 1098 XG Amsterdam, Netherlands}

\forColl{KM3NeT} 

\emailAdd{tjuanve@nikhef.nl}
\emailAdd{jseneca@nikhef.nl}

\abstract{The discovery of a high-energy cosmic neutrino flux has paved the way for the field of neutrino astronomy. For a large part of the flux, the sources remain unidentified. The KM3NeT detector, which is under construction in the Mediterranean sea, is designed to determine their origin. KM3NeT will instrument a cubic kilometre of seawater with photomultiplier tubes that detect Cherenkov radiation from neutrino interaction products with nanosecond precision. For single cascade event signatures, KM3NeT already showed that it can reach degree-level resolutions, greatly increasing the use of these neutrinos for astronomy. In this contribution, we further refine the cascade reconstruction by making a more detailed model of the neutrinos events and including additional information on the hit times. The arrival time of light can be used to improve the identification of double cascade signatures from tau neutrinos, and the angular resolution of both single and double cascade signatures. Sub-degree resolution is achieved in both cases.}
\FullConference{37th International Cosmic Ray Conference (ICRC 2021)\\
July 12th – 23rd, 2021 \\
Online – Berlin, Germany}

\maketitle

\section{Introduction}

\subsection{Neutrino astronomy}

Neutrino properties make it an excellent complement to other cosmic messengers. The neutrality of the neutrino prevents deflections due to ambient magnetic fields when travelling to the Earth. Further, very low cross-sections enable the study of particle accelerators closer to the source, as well as preventing attenuation of the neutrino flux on the way to the Earth. The probing power of the neutrino as a cosmic messenger and implications for astroparticle physics and the understanding of cosmic ray production mechanisms and sources has been extensively documented \cite{becker, learned}.

A cosmic neutrino flux has been discovered by the IceCube detector \cite{ic-flux}, and while observations hinting at neutrino origins have been made \cite{blazar}, the sources of cosmic neutrinos remain elusive. 

To enable the discovery of neutrino sources, a good angular resolution of the detector is crucial. For this reason, the track signatures caused by $\nu_\mu$ charged current interactions have been of the highest interest for searching for neutrino sources, IceCube achieving a resolution of $0.25^{\circ}$ \cite{ic-source}, and below $0.1^{\circ}$ for KM3NeT/ARCA \cite{recokm3net}. However, about two thirds of cosmic neutrinos are expected to produce a signature that is not a track, while currently only 20\% of observed IceCube events have been tracks \cite{ic-events}. A good angular resolution on cascade events therefore represents an encouraging opportunity for neutrino source discovery.
For cascades, IceCube achieves a resolution of $10^{\circ}$ at best \cite{ic-cascade}, while the current high energy KM3NeT cascade reconstruction algorithm achieves a resolution of $1.5^{\circ}$ at best \cite{loi}. This work aims at improving further on the cascade signature reconstruction, and implementing the reconstruction for the most common $\nu_\tau$ charged current signature.

\subsection{KM3NeT/ARCA}

KM3NeT is two neutrino detectors currently under construction on the bottom of the Mediterranean sea. ORCA is the component focusing on oscillation studies of atmospheric neutrinos travelling through the Earth. ARCA is the high energy neutrino telescope complement and subject of this work, located in Portopalo di Capo Passero in Sicily, Italy. The detector consists of a 3-D grid of digital optical modules (DOMs) that each contain 31 photomultiplier tubes (PMTs) \cite{bruijn}. 
The modules are mounted to vertical detection units (DUs) containing 18 DOMs spaced about 38 meters apart. The DUs are mounted to the seafloor and grouped in building blocks of 115 DUs with an average horizontal spacing of 95 meters. The complete ARCA detector will consist of two building blocks \cite{loi}.




\subsection{Direction reconstruction of cascades}

The two main neutrino event signatures detected by KM3NeT are \emph{tracks} and \emph{cascades}. Hadronic cascades are produced at the vertex of all charged-current (CC) and neutral-current (NC) neutrino interactions. At high energies, the neutrino interacts through deep inelastic scattering, kicking out an energetic quark which forms the hadronic cascade. For CC interactions, an additional lepton corresponding to the neutrino flavour is produced. For electrons, this leads to an electromagnetic cascade, and thus a cascade signature, while muons travel through large parts of the detector. Highly-relativistic muons can travel several kilometers through seawater while producing Cherenkov light resulting in an extended track signature when comparing with other events. Muons produced by cosmic rays interacting in the atmosphere also reach and travel through the detector. The tau has a mean lifetime of $2.903 \pm 0.005 \times 10^{-13}$ s and decays into hadrons and leptons \cite{tau}. The branching ratio to an electron or hadrons is 0.8261 which can result in a \emph{double cascade} signature for high enough $\tau$-decay lengths. The cascades are separated by an average $5 \frac{ \text{cm}}{\text{ TeV}}$ due to time dilatation. The double cascade signature distinguishes itself from regular cascades when the tau length is more than a few meters, which is the typical particle cascade size at TeV energies. 

The direction reconstruction of the track signature utilizes the detection time of Cherenkov light on PMTs near the track in order to fit the direction of the muon. The standard KM3NeT cascade reconstruction does not use the detection times to reconstruct the direction, but uses the presence or absence of hits on each PMT as a measure of the light intensity. This work presents two new reconstruction algorithms for single and double cascades where we include the timing information for direction reconstruction and improve on the modelling of cascades by taking into account their elongated shape. This effort is motivated by the idea that an improved description of the events and additional information leads to an improvement in reconstruction performance. The improved model with timing information is also expected to improve due to the \textit{lever arm effect} which the track reconstruction also benefits from. The early and late parts of both single and double cascade events are strongly restricted in position thanks to the arrival time of the light that they produce. This places the start and end of the event along the direction of the event, achieving a better angular resolution. The method is described first in \ref{sec:method}, followed by the performance for single and double cascades in \ref{sec:performance}. We conclude with a summary and short outlook in \ref{sec:summary}.



\section{Method}
\label{sec:method}

The standard KM3NeT single cascade reconstruction (Aashowerfit) fits the spatial Cherenkov profile to the PMT and hit positions to estimate the direction and energy of a neutrino. The vertex and time of the event are estimated by minimizing the hit time residuals assuming isotropic light emission from the shower maximum. 

The likelihood function is built from the Poisson probability of no hits occurring. If a hit did occur on the PMT, the complementary probability is used. The likelihood is then the product of the Poisson probabilities for each PMT hit or no-hit status,
\begin{equation}
    \mathcal{L}_{\text{hit/no-hit}} = \prod_{ \text{hit PMTs} } 1 - e^{-N_s - N_{bg}} \prod_{ \text{no-hit PMTs}} e^{-N_s - N_{bg}}
\label{eq:hitnohitlik}
\end{equation}
where $N_s$ is the expected number of photo-electrons (n.p.e.) due to a cascade hypothesis and $N_{bg}$ is the expected background n.p.e. $N_{bg}$ is obtained by fitting ARCA data to a constant $^{40}K$ background hypothesis. For Aashowerfit, $N_s$ is obtained from a model built from interpolated Monte Carlo simulations of 1 PeV $\nu_e$ CC events. In the new reconstruction, $N_s$ is obtained from a semi-analytical Cherenkov light model built with interpolated simulations of electromagnetic cascade energy deposition and light emission profiles. Aashowerfit is used as a prefit for the single and double cascade reconstructions using timing information.

\subsection{First hit information}

\begin{figure}[h]
\centering
\includegraphics[width=0.4\textwidth]{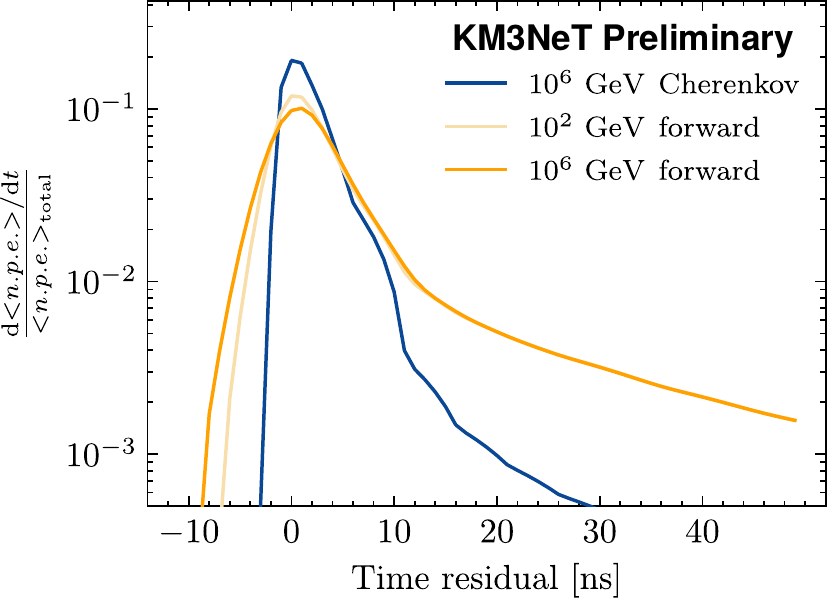}
\caption{Time residuals for a $10^2$ and $10^6$ GeV single cascade, for a PMT in the cherenkov cone, and in the forward direction. Note that different profiles are observed depending on where the event is pointing with respect to the PMT.}
\label{fig:1sthit}
\end{figure}

The hit/no-hit likelihood from equation \ref{eq:hitnohitlik} is the sum of the probabilities for PMTs to have seen light or not during a cascade event. This likelihood can be extended with the information of when each hit PMT was hit. The analogue signals from the PMTs are digitized inside the DOMs and return a \emph{time}, at which the signal surpasses a threshold, and a \emph{time-over-threshold} (ToT). Photons on the same PMT that are closely spaced in time result in a single hit with the time of the first hit. Only the hit time of the first hit on every PMT is used when including timing information. Figure \ref{fig:1sthit} shows the time residuals for a single cascade.

The hit time information is crucial in the reconstruction of track-like events induced by muons. A highly-relativistic muon travels long distances under water while producing Cherenkov light. The Cherenkov wavefront is the first light hitting a PMT and this information is used to fit the direction of the muon, also benefiting from the lever-arm effect. For cascades, Aashowerfit utilizes the timing information to fit the vertex position and time, but ignores the time in the direction reconstruction.

\subsection{Fit routine}

The likelihood can be re-defined to include a time component 
\begin{equation}
    \mathcal{L}_{\text{time}}= \prod_{ \text{1st hits} } P_{\text{1st}}(t)
\label{eq:timelik}
\end{equation}
where $P_{\text{1st}}(t)$ is the probability density for the first hit to occur at time $t$ given that a hit occurs. The full likelihood can then be written as
\begin{equation}\label{eq:likfunc}
    \mathcal{L} = \prod_{ \text{1st hits} } P_{\text{1st}}(t) \prod_{ \text{hit PMTs} } 1 - e^{-N_s - N_{bg}} \prod_{ \text{no-hit PMTs}} e^{-N_s - N_{bg}}
\end{equation}
using both the hit/no-hit information and the timing information. The reconstruction is a maximum likelihood estimator of the likelihood in eq. \ref{eq:likfunc}.

\subsubsection{Single cascades}

The single cascade algorithm using timing information consists of three steps:
\begin{enumerate}
    \item Aashowerfit prefit. 
    \item Precision position and time prefit.
    \item Full likelihood maximisation using single cascade hypothesis.
\end{enumerate}

The timing information of the first hits is sensitive to the position and time of vertex of the cascade. A good vertex position and time prefit is therefore constructed using the timing of the first hits only. The Aashowerfit vertex time is used as a starting point for the prefit. For the position, 13 positions are chosen. The Aashowerfit vertex position, and 12 positions equally spaced on a sphere of radius 1 m, centered on the Aashowerfit vertex position. Starting from each position, the first hits only are used to reconstruct the position and time of the vertex, using the first hit likelihood and elongated model. Out of the 13 results, the track with the highest likelihood is picked as the prefit result. This process is re-iterated once with the new best-fit vertex to further improve the prefit and kick outliers out of local minima. The full likelihood, eq. \ref{eq:likfunc}, is then maximized with the non-hit PMTs, and the time of the hit PMTs.

\subsubsection{Double cascades}

The double cascade algorithm consists of three steps:
\begin{enumerate}
    \item Aashowerfit prefit
    \item Tau length prefit
    \item Time likelihood maximisation using double cascade hypothesis.
\end{enumerate}
The Aashowerfit prefit assumes a single cascade and estimates a vertex somewhere in between or in front of the first and second cascade. The tau length prefit searches along the Aashowerfit direction to find the position of the first and second cascade. This results in a first estimate of the travelling length of the tau lepton. The results of the prefits are used as starting values in the final fit where the time likelihood (equation \ref{eq:timelik}) is maximised with respect to the vertex of the first cascade, the direction of both cascades, the length of the tau and the energy division of the first and second cascade. The high-energy cascades are assumed to be colinear and separated by the speed of light.

\newpage
\section{Performance}
\label{sec:performance}

\subsection{ Single cascades }

\begin{figure}[h]%
    \centering
    \subfloat[\centering  ]{{\includegraphics[width=7cm]{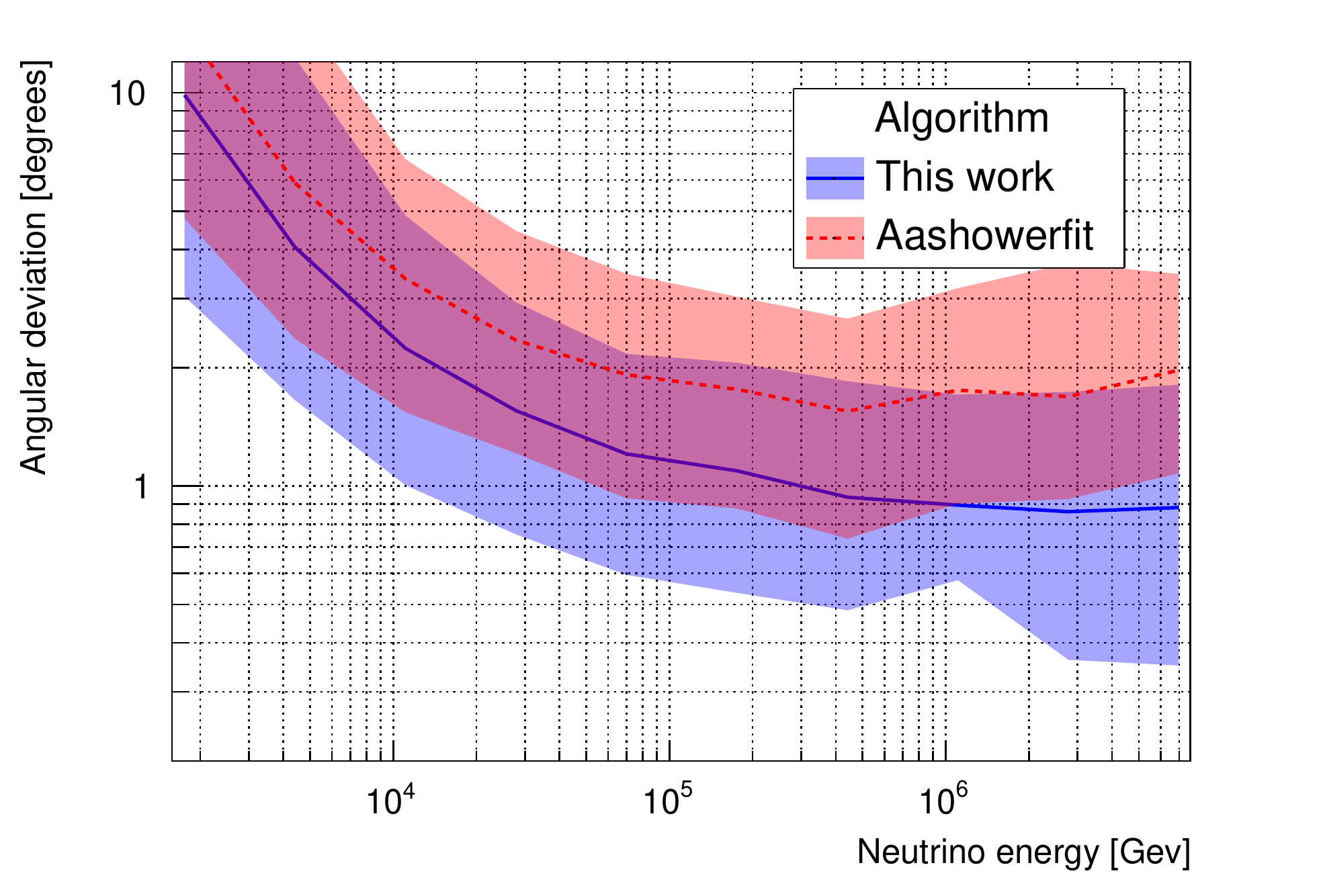}}}%
    \qquad
    \subfloat[\centering ]{{\includegraphics[width=7cm]{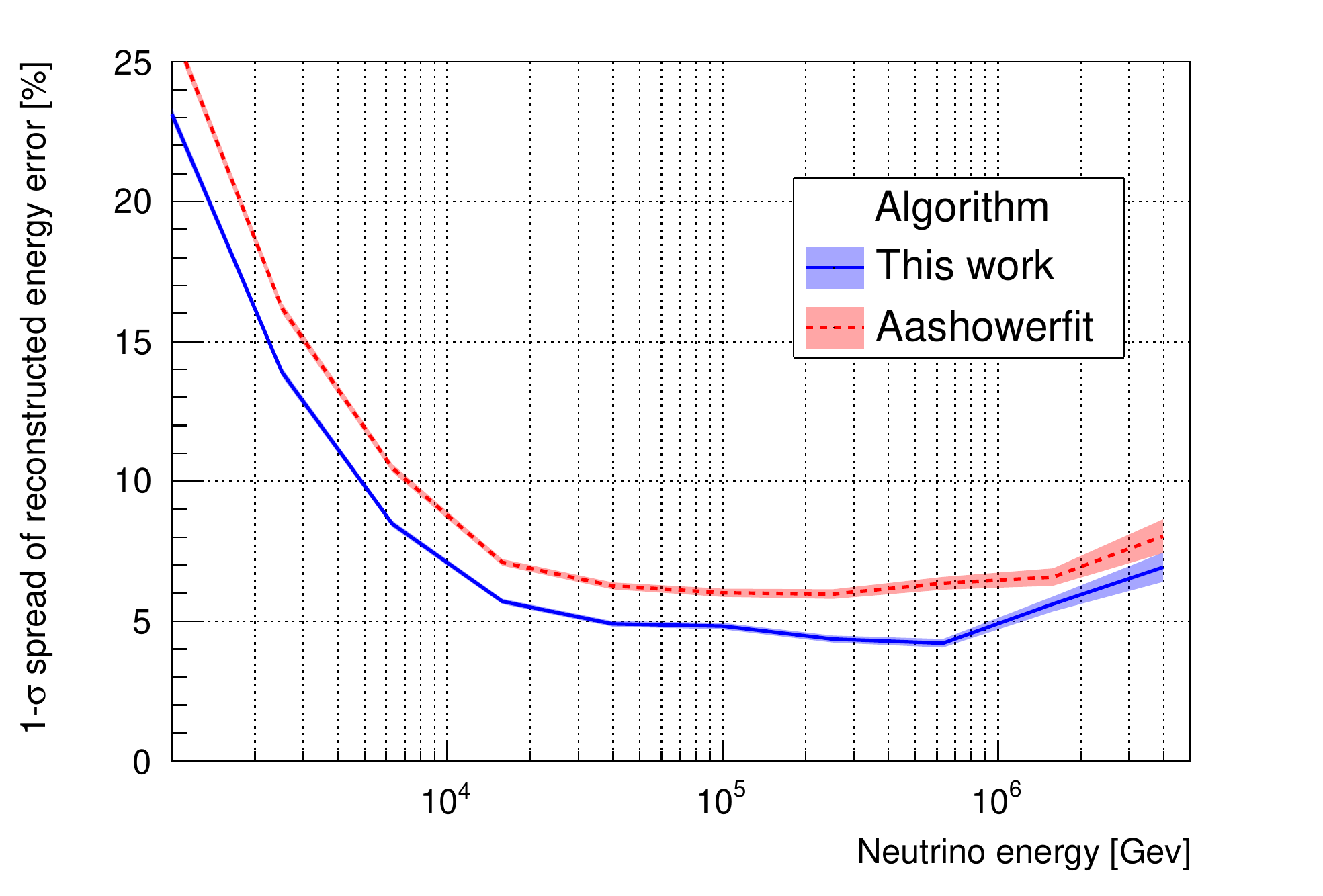} }}%
    \caption{Direction \textbf{(a)} and energy \textbf{(b)} reconstruction performance for single cascade reconstruction.}
    \label{fig:res_elec}%
\end{figure}

Figure \ref{fig:res_elec} shows the median angular deviation and spread of relative energy difference for the single cascade reconstruction. There is a clear improvement in the median when including timing information. The spread of relative energy difference between reconstructed energy and true energy also shows an improvement.

\begin{figure}[h]%
    \centering
    \subfloat[\centering  ]{{\includegraphics[width=7cm]{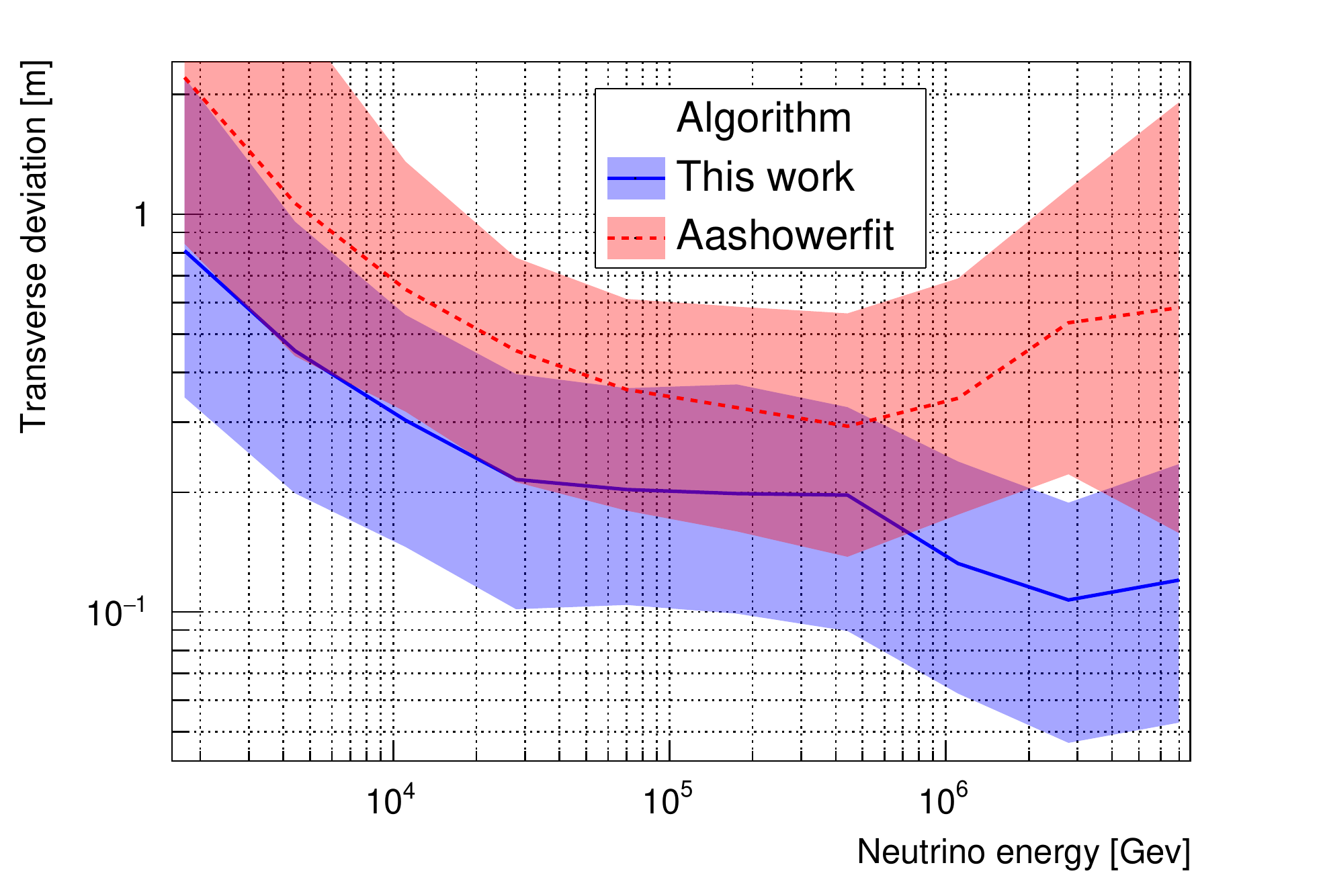}}}%
    \qquad
    \subfloat[\centering ]{{\includegraphics[width=7cm]{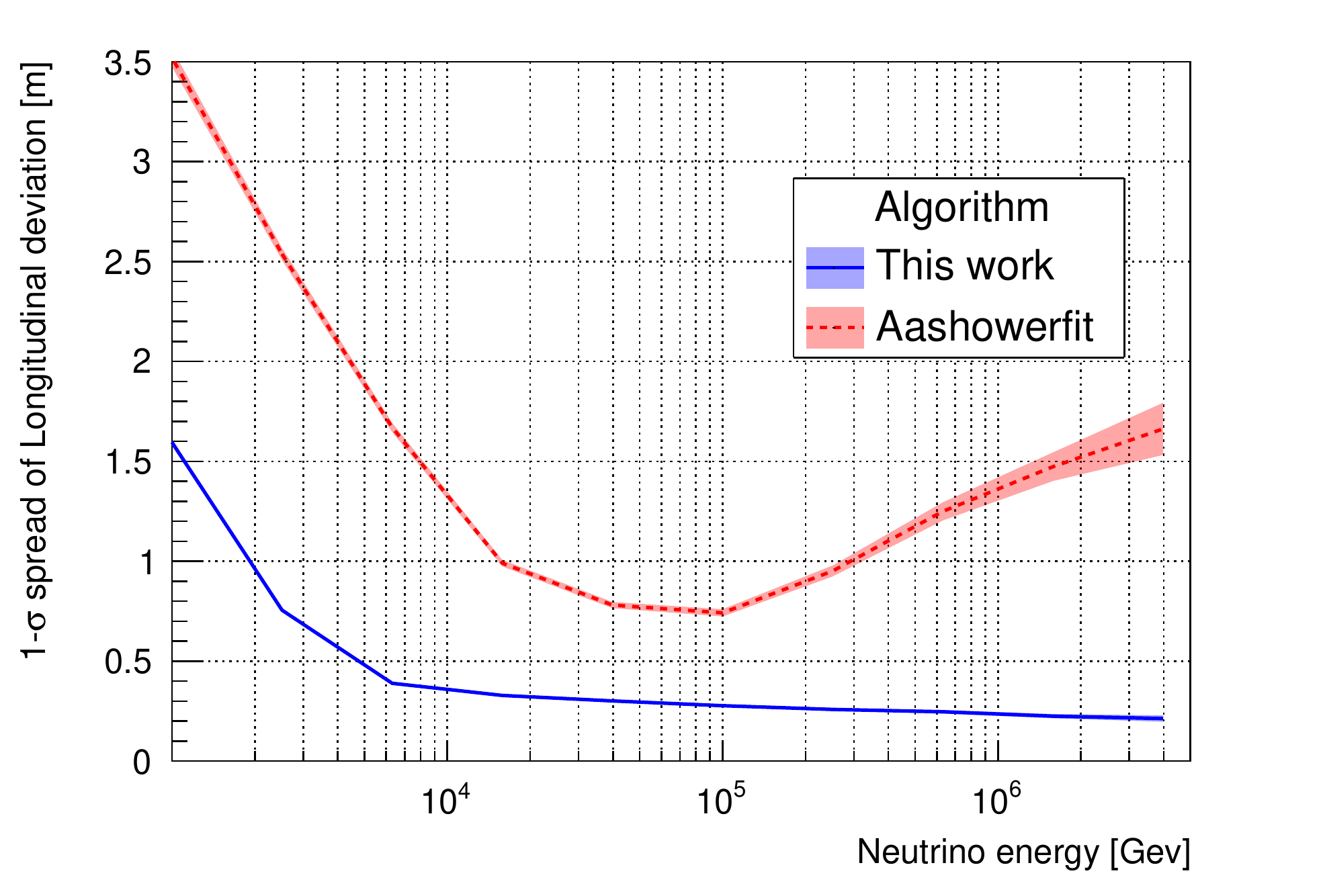} }}%
    \caption{Transverse \textbf{(a)} and longitudinal \textbf{(b)} position deviation for single cascade reconstruction.}
    \label{fig:posres_elec}%
\end{figure}

Figure \ref{fig:posres_elec} shows the position deviation for the single cascade reconstruction. The position resolution improves when including timing information in the elongated cascade model for both the transverse and longitudinal plane.

\newpage
\subsection{ Double cascades }

\begin{figure}[h]
\centering
\includegraphics[width=0.7\textwidth]{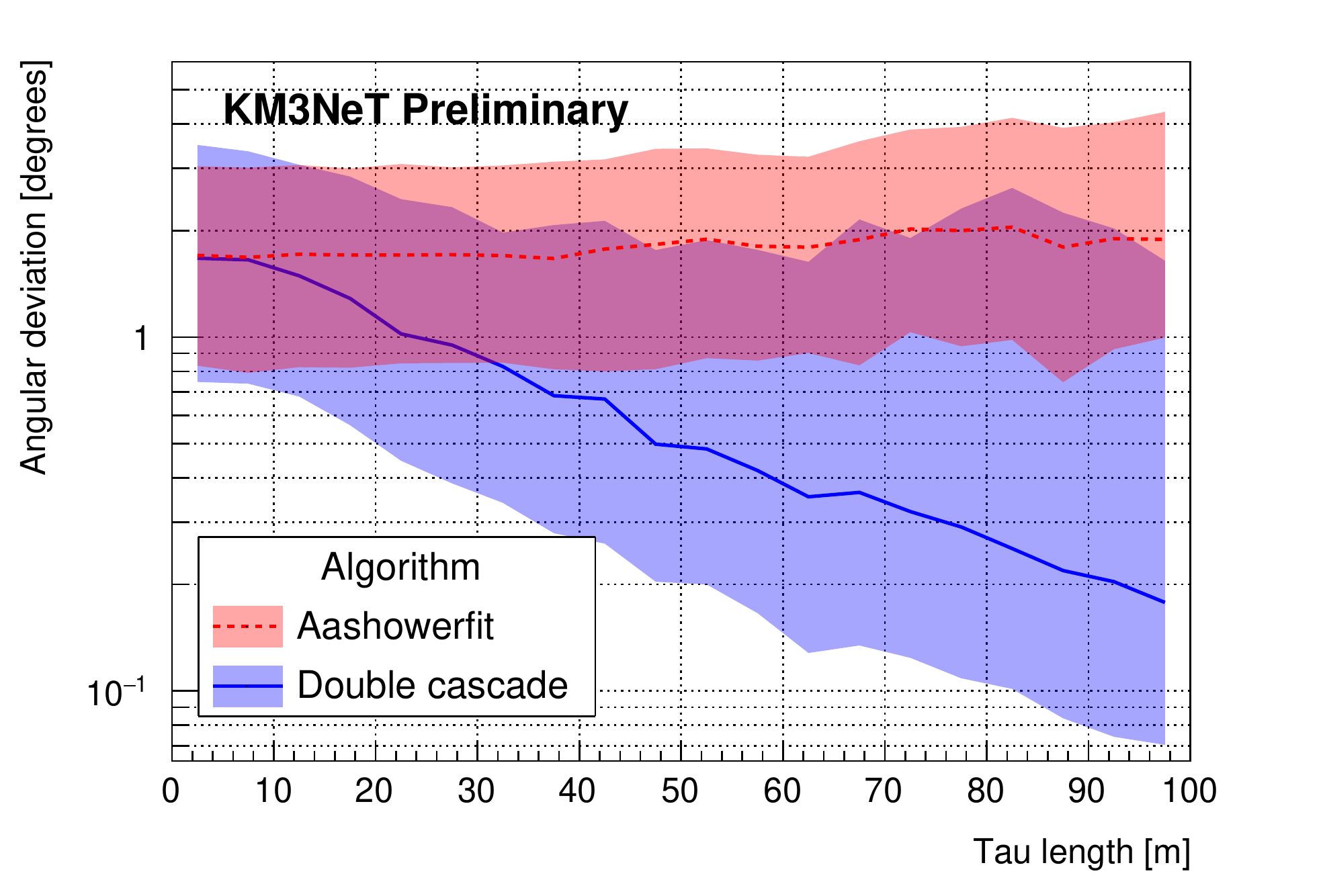}
\caption{Angular deviation as a function of the true tau length.}
\label{fig:angres_tau}
\end{figure}

Figure \ref{fig:angres_tau} shows the angular deviation between the reconstructed direction and true direction for double cascade events. The events are weighted with a $1.2 \cdot 10^{-8} \cdot E^{-2} \text{ GeV }^{-1}\text{sr }^{-1}\text{s }^{-1}\text{cm}^{-2}$ spectrum and there has been an event selection based on the Aashowerfit output. The reconstructed vertex is required to be within the inner half volume of the detector and the reconstructed energy above 100 TeV.

\begin{figure}[h]%
    \centering
    \subfloat[\centering  ]{{\includegraphics[width=7cm]{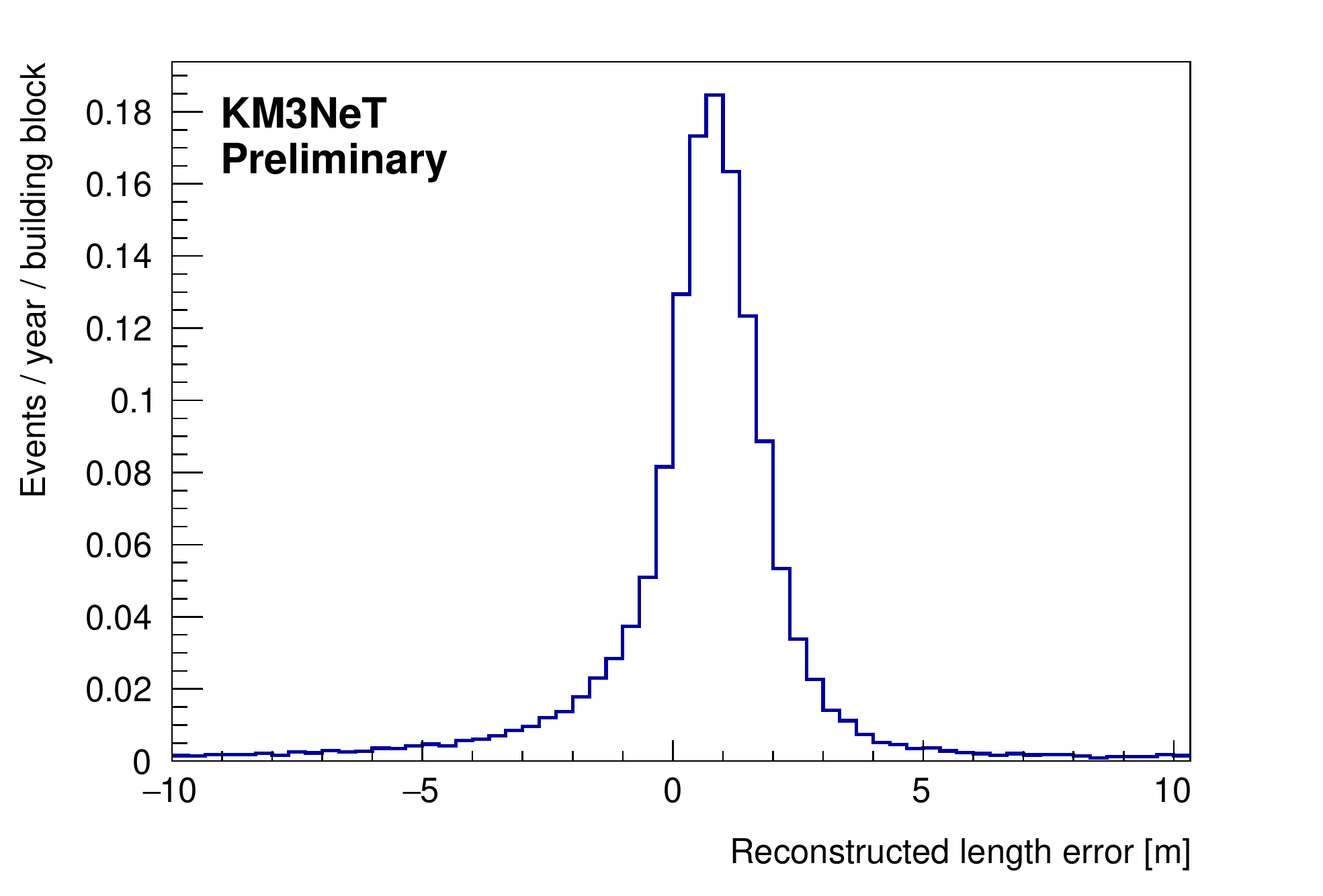}}}%
    \qquad
    \subfloat[\centering ]{{\includegraphics[width=7cm]{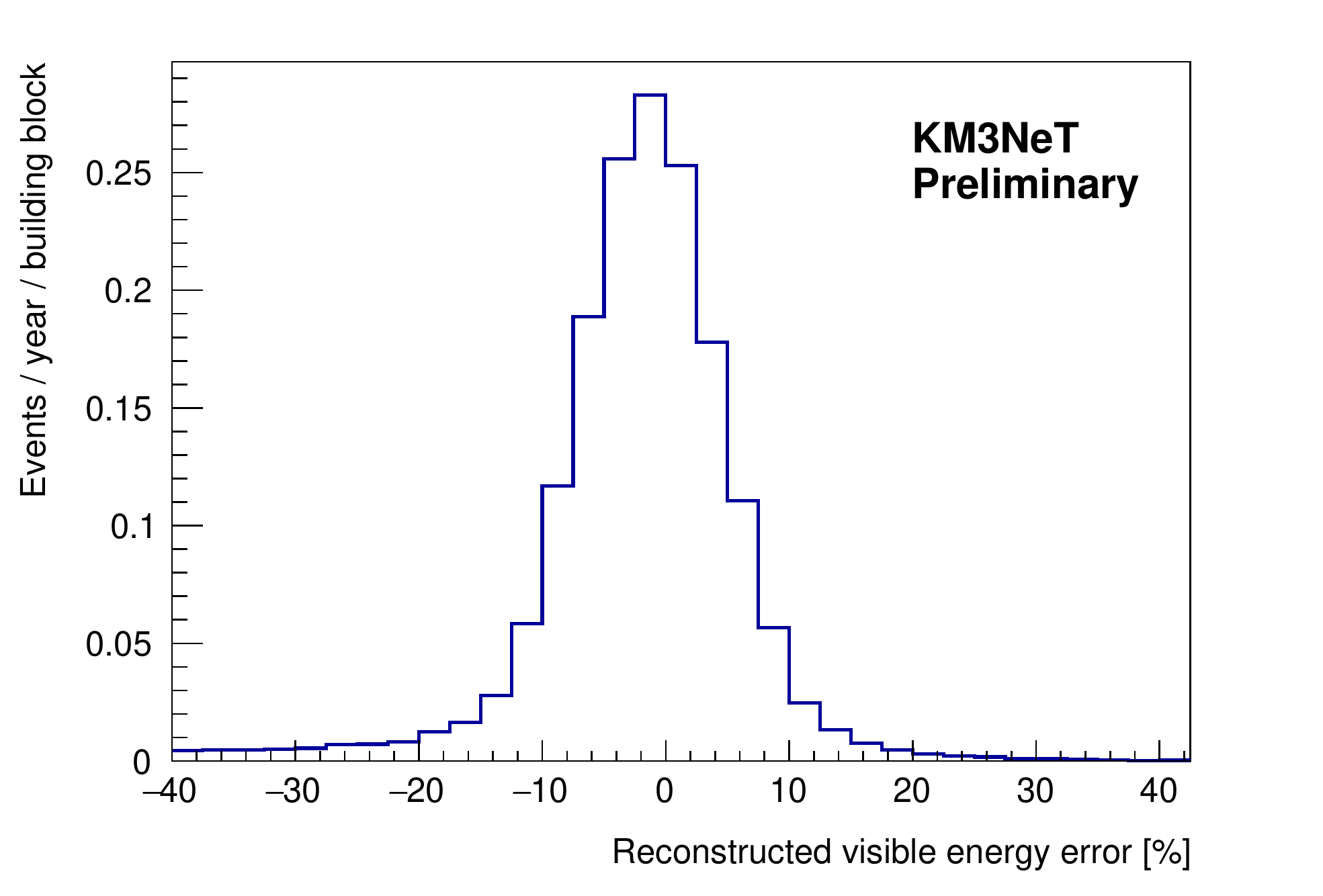} }}%
    \caption{Reconstructed length \textbf{(a)} and energy error \textbf{(b)} for the double cascade reconstruction. The median and 68\% quantiles of the reconstructed length error are $0.72^{+1.23}_{-1.95}$ meters. The median and 68\% quantiles of the reconstructed energy error are $-1.75^{+6.11}_{-6.90}$\%.}
    \label{fig:lendifferes}%
\end{figure}

Figure \ref{fig:lendifferes} shows the reconstructed length error and the reconstructed visible energy error.

\section{Summary}
\label{sec:summary}

We have a presented two new reconstruction algorithms which use the timing information and elongation emission profile of cascades. The median angular deviation improves over the whole energy range and drops below 1 degree for single cascades of 300 TeV. We also see an improvement in the position resolution that benefits the double cascade reconstruction. For double cascades, the angular deviation drops below 1 degree for tau lengths longer than 25 meters. The improvement with respect to single cascades comes from a stronger lever-arm effect due to the large extension of double cascade events. We find a reconstructed length error spread of 3.17 meters and a reconstructed visible energy error spread of 13\%.

\clearpage
\section*{Full Author List: KM3NeT Collaboration}

\scriptsize
\noindent
M.~Ageron$^{1}$,
S.~Aiello$^{2}$,
A.~Albert$^{3,55}$,
M.~Alshamsi$^{4}$,
S. Alves Garre$^{5}$,
Z.~Aly$^{1}$,
A. Ambrosone$^{6,7}$,
F.~Ameli$^{8}$,
M.~Andre$^{9}$,
G.~Androulakis$^{10}$
% ...
\\

\noindent
$^{1}$Aix~Marseille~Univ,~CNRS/IN2P3,~CPPM,~Marseille,~France. \\
$^{2}$INFN, Sezione di Catania, Via Santa Sofia 64, Catania, 95123 Italy. \\
$^{3}$Universit{\'e}~de~Strasbourg,~CNRS,~IPHC~UMR~7178,~F-67000~Strasbourg,~France. \\
$^{4}$Universit{\'e} de Paris, CNRS, Astroparticule et Cosmologie, F-75013 Paris, France. \\
$^{5}$IFIC - Instituto de F{\'\i}sica Corpuscular (CSIC - Universitat de Val{\`e}ncia), c/Catedr{\'a}tico Jos{\'e} Beltr{\'a}n, 2, 46980 Paterna, Valencia, Spain. \\
$^{6}$INFN, Sezione di Napoli, Complesso Universitario di Monte S. Angelo, Via Cintia ed. G, Napoli, 80126 Italy. \\
$^{7}$Universit{\`a} di Napoli ``Federico II'', Dip. Scienze Fisiche ``E. Pancini'', Complesso Universitario di Monte S. Angelo, Via Cintia ed. G, Napoli, 80126 Italy. \\
$^{8}$INFN, Sezione di Roma, Piazzale Aldo Moro 2, Roma, 00185 Italy. \\
$^{9}$Universitat Polit{\`e}cnica de Catalunya, Laboratori d'Aplicacions Bioac{\'u}stiques, Centre Tecnol{\`o}gic de Vilanova i la Geltr{\'u}, Avda. Rambla Exposici{\'o}, s/n, Vilanova i la Geltr{\'u}, 08800 Spain. \\
$^{10}$NCSR Demokritos, Institute of Nuclear and Particle Physics, Ag. Paraskevi Attikis, Athens, 15310 Greece. \\
% ...
$^{55}$Universit{\'e} de Haute Alsace, 68100 Mulhouse Cedex, France. \\
% ...

\end{document}